\documentclass[12pt,final,a4paper,notitlepage]{iopart}
\usepackage{graphicx}
\usepackage{bm}
\usepackage{hyperref}
\usepackage{color,array}
\begin{document}


\title{Stochastic models of multi-channel particulate transport with blockage}
\author{Chlo\'e Barr\'e, Gregory Page,  Julian Talbot and Pascal Viot}
\address{
Laboratoire de Physique Th\'eorique de la Mati\`ere Condens\'ee,
Sorbonne Universit\'e, CNRS  UMR 7600, 4, place Jussieu, 75252 Paris Cedex 05, France
}
\begin{abstract}

Networks of channels conveying particles are often subject to blockages due to the limited carrying capacity of the individual channels. If the channels are coupled, blockage of one causes an increase in the flux entering the remaining open channels leading to a cascade of failures. Once all channels are blocked no additional particle can enter the system.  If the blockages are of finite duration, however,  the system reaches a steady state with an exiting flux that is reduced compared to the incoming one. We propose a stochastic model consisting of $N_c$ channels each with a blocking threshold of $N$ particles.  Particles enter the system according to a Poisson process with the entering flux of intensity $\Lambda$ equally distributed over the open channels. Any particle in an open channel exits at a rate $\mu$ and a blocked channel unblocks at a rate $\mu^*$.  We present a method to obtain the exiting flux in the steady state, and other properties, for arbitrary $N_c$ and $N$ and we present explicit solutions for $N_c=2,3$. We apply these results to compare the efficiency of conveying a particulate stream of intensity $\Lambda$ using different channel configurations. We compare a single ``robust" channel with a large capacity with multiple ``fragile" channels with a proportionately reduced capacity.  The ``robust" channel is more efficient at low intensity, while multiple, ``fragile" channels have a higher throughput at large intensity. We also compare $N_c$ coupled channels with $N_c$ independent channels, both with threshold $N=2$. For $N_c=2$ if $\mu^*/\mu>1/4$, the coupled channels are always more efficient. Otherwise the independent channels are more efficient for sufficiently large $\Lambda$. 
  
\end{abstract}
\jl{3}
\submitted{}

\date{\today}

\maketitle

\section{Introduction}


Particulate flow through a network of narrow channels may be subject to blockage due to the limited carrying capacity of 
the individual channels. Blockage of one channel results in an increased load on the remaining open channels that can 
trigger a cascade of additional blockages, ultimately leading to a complete breakdown of the system \cite{Watts2002,Crucitti2004,Zhao2015,Kim2005}. 
If the blockage is of finite duration, on the other hand, the system will eventually reach a steady state.

An analogous phenomenon of multiple failures can be observed in textile fibers when an external force is exerted.
The fiber bundle model (FBM) \cite{Peirce1926,daniels1945,coleman1956} 
consists of a number of parallel threads subjected  to an applied load. 
If the load on a single thread exceeds its threshold, the thread breaks and the global load is then redistributed over the remaining intact
threads\cite{pradhan2010}.  
Blackouts in power distribution networks are generally preceded by a cascade of failures that results from
local overloads\cite{Dobson2005}. Earthquakes\cite{DidierSornette1992,Newman1994}, vehicular traffic jams \cite{Chakrabarti2006}, network traffic jams\cite{Ezaki2015,Daganzo2011,Gayah2011,Ji2012}, material fractures\cite{Mishnaevsky2009,Raischel2006} and internet attacks (DoS) \cite{bhunia2014} exhibit similar features.

We focus here on particulate flow in channels  with a limited carrying capacity defined by a fixed number of particles, $N$.
When this threshold is reached, a blockage occurs and no additional particle can enter the channel for the duration of the blockage. For a purely ballistic transport in the channel, Gabrielli et al. \cite{Gabrielli2013} introduced non-Markovian models  in which particles randomly  enter a channel and traverse it at a constant velocity.
For permanent blockage,  exact solutions can be obtained\cite{Gabrielli2013,Talbot2015} for small thresholds ($N=2,3$), but only numerical results are available for larger values of $N$. Several extensions of these models have been considered, including a time-dependent incoming flux\cite{Barre2015}, temporary blockage\cite{BTV2013}, as well as multiple channel transport where  cascades of failures were observed\cite{Barre2015b}. However, it becomes increasingly difficult to obtain exact solutions as the model complexity is increased.

In this article we propose a class of Markovian  models, inspired by the queuing theory\cite{Adan2002,Medhi2003},  that encompasses all previous situations: temporarily or permanent blockage, and  single or multi-channel systems for which stationary and time-dependent solutions can be obtained. These results provide an efficient tool for optimizing the design particle transport systems.

\begin{figure}[t]
\begin{center}
\includegraphics[width=10cm]{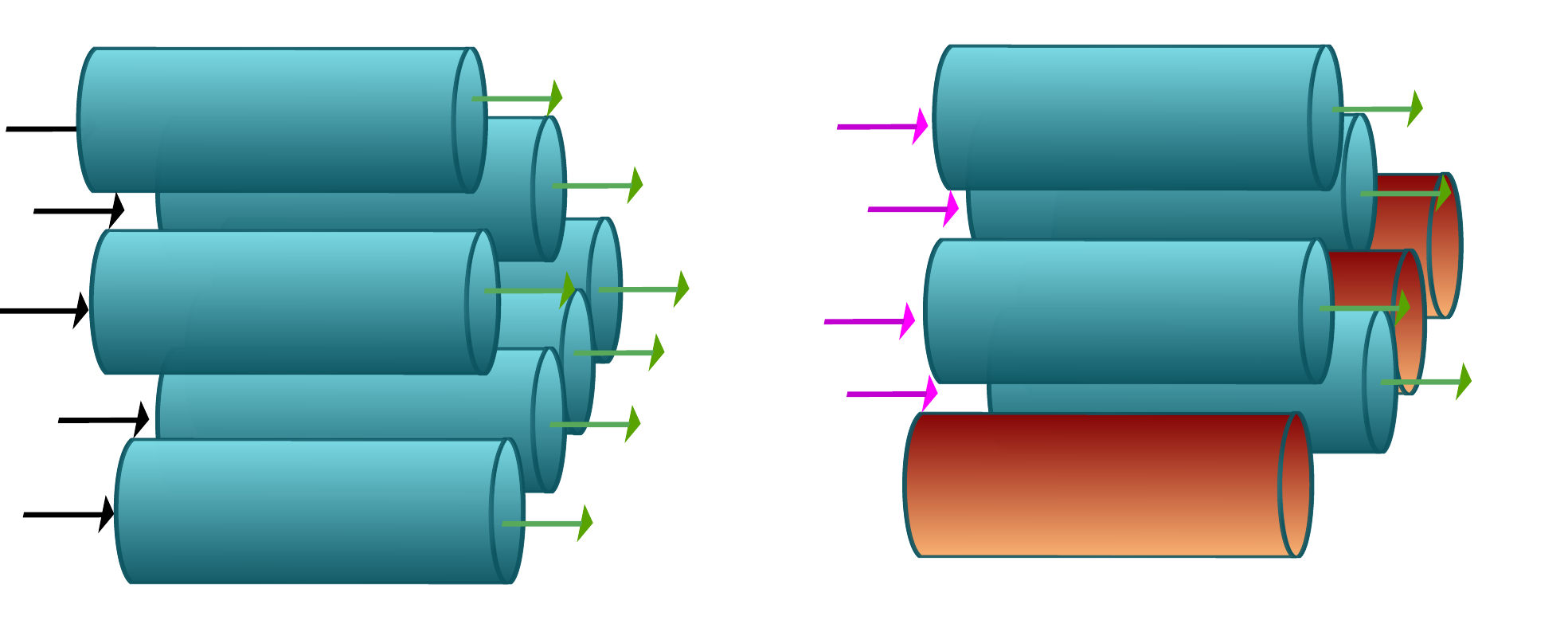}
\end{center}
\caption{(Color online) The channel network is composed of $N_c$ channels. A total entering flux $\Lambda=N_c\lambda$ is evenly distributed
over the channels. An individual channel has a maximum carrying capacity: if $N$  particles  are simultaneously present the channel a blockage occurs. If a channel is open any particle that it contains exits at a rate $\mu$ (green arrows). The channels are coupled such that the constant total flux is redistributed 
over the open channels: (left) all channels are open, the channel flux is equal to $\lambda$ (black arrows); (right) three channels are blocked, the channel becomes equal to $\lambda'=7\lambda/4$ (mauve arrows).
 }\label{fig:cylin}
\end{figure}

\section{Model}
We consider a system composed of $N_c$ channels. 
A channel is open if the number of particles simultaneously present inside  is less than $N$ and blocked when the threshold of $N$ particles is reached. In the latter case, no more particles can enter  until the channel is flushed. 
Particles randomly enter an open channel according a Poisson process with intensity (rate) $\Lambda/(N_c-k)$ where $k$ is the number of 
blocked channels at time $t$.
Particles randomly exit an open channel after an exponentially distributed time with rate $\mu$ and independently of the number of particles in the channel. A blocked channel  reopens, and releases $N$ blocked particles, after an exponentially distributed time with rate $\mu^*$.  
A channel can be in one of $N+1$ states, corresponding to an index ranging from $0$ to $N$. 
When $i=N$, the channel is blocked. For $N_c$ channels, the time evolution of the system can be described by introducing the  probabilities 
associated with the number of channels in each state,
$\pi (i_N,i_{N-1},...,i_1)$, where each index $i_j$ goes from $0$ to $N_c$, with the constraint $\sum_{j=0}^{N}i_j=N_c$. $i_0$ corresponds to 
the number of empty channels. The evolution of the  channel state depends only on the network state at time $t$ and hence the process is  Markovian.
However, as we will see below, the number of probabilities  increases roughly  as $N_c^N$, which leads to some cumbersome calculations in general.
In the following section we focus on single channel models for general threshold number $N$. In section \ref{sec:multiple}, we consider $N_c$ channels with a fixed threshold $N=2$.

\section{Single channel models}

Particles enter the channel according to a Poisson process with a constant intensity $\lambda$. 
Particle egress from the channel is also modeled as Poisson process with  constant rate $\mu$. When $N$ particles are in the channel at the same time, a blockage occurs with the consequence that all incoming 
particles are rejected until the channel reopens after a time given by an exponentially distributed blockage time with rate $\mu^*$. When the channel  reopens, 
all trapped particles are ejected at the same time.

For single channel models, the stochastic dynamics is described  by using $N+1$ state probabilities, $\pi(0,0,t), \pi(0,1,t),\cdots, \pi(N-1,0,t)$   and $\pi(1,0,t)$ which corresponds 
to $0,1,\cdots,N-1$ and $N$, particles in the channel, respectively. If $P(t)=(\pi(0,0,t), \pi(0,1,t),\cdots, \pi(N-1,0,t),\pi(1,0,t))$ denotes the state vector, the time evolution of the process is described by
\begin{equation}\label{eq:scm}
\frac{dP(t)}{dt}=P(t).Q_{N+1} 
\end{equation}
where $Q_{N+1}$ is the $(N+1)\times (N+1)$ matrix,
\begin{equation}
 Q_{N+1}=\left(
 \begin{array}{lllllll}
\scriptstyle -\lambda&\scriptstyle \lambda&0&...&0&0\\
 \scriptstyle\mu &\scriptstyle-(\lambda+\mu)&\lambda&...&0&0\\
 \vdots&\vdots&\vdots&\ddots&\vdots&\vdots\\
 0&...&\scriptstyle\mu&\scriptstyle-(\lambda+\mu)&\lambda&0\\
 0&...&0&\scriptstyle\mu&\scriptstyle-(\lambda+\mu)&\lambda\\
 \mu^*&...&...&0&0&-\scriptstyle\mu^*
\end{array}
\right)
\end{equation}
that can be interpreted as follows. Except for the first and  last column, the change of  state $i$ has two gain terms and two loss terms: The two loss terms correspond to  a channel with $i$ particles (with $0<i<N$) where a particle enters at time $t$  (with a rate $\lambda$) or where a
particle exits (at rate $\mu$). The two gain terms correspond to the entrance of a particle (with a rate $\lambda$) in a channel with $i-1$ particles, and to the exit of a particle (at rate $\mu$) from a channel with $i+1$ particles.
The description is completed  by considering the two boundary situations: for an empty channel, there is one loss term corresponding 
to a particle entrance and two gain terms: the first corresponds to a particle exit from an 
empty channel and the second to the release of a blocked channel (with $N$ particles). Conversely, for a blocked channel, 
one has a single loss term corresponding to a release with a rate $\mu^*$ and a gain term corresponding to a particle entering a channel containing $N-1$ particles.

The time evolution of $P(t)$ is  supplemented by the conservation of the total probability, $\sum_{i=0}^{N_1} \pi(0,i,t)+\pi
(1,0,t)=1$. Consequently, the sum of each row of the transition matrix
is obviously equal to $0$, which leads to a zero eigenvalue of the matrix.

The throughput of the channel is given by
\begin{equation}
 j(t)=\mu\sum_{k=1}^{N-1}\pi(0,k,t)+N\mu^*\pi(1,0,t),
\end{equation}
which accounts for the exit of one particle at time $t$ whatever the state of the open channel and the release of a blocked channel where $N$ particles exit at the same time.

In the stationary state, $j$ can be expressed as
\begin{equation}\label{eq:fluxs}
 j=\lambda(1-\pi(1,0)),
\end{equation}
i.e., the incoming flux times the probability that the channel is open.

\begin{figure}[t!]
\begin{center}
\includegraphics[width=7.0cm]{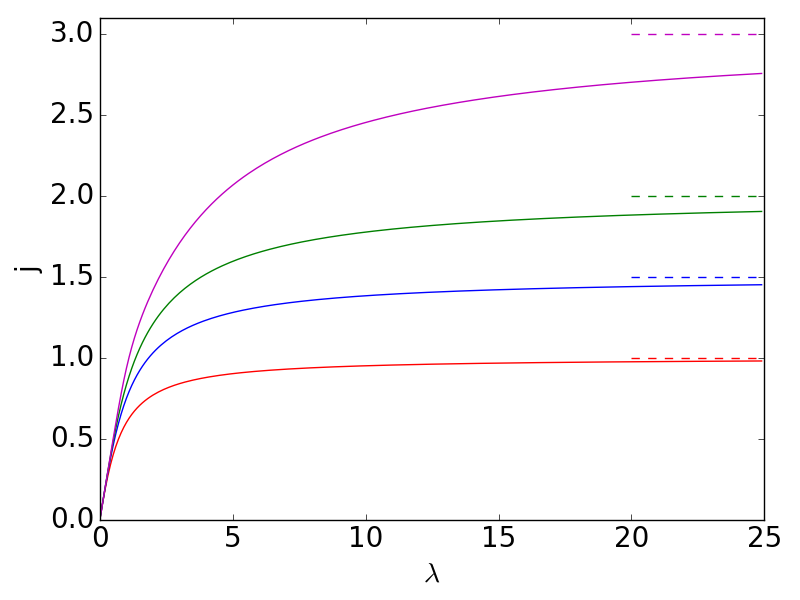}\includegraphics[width=7.0cm]{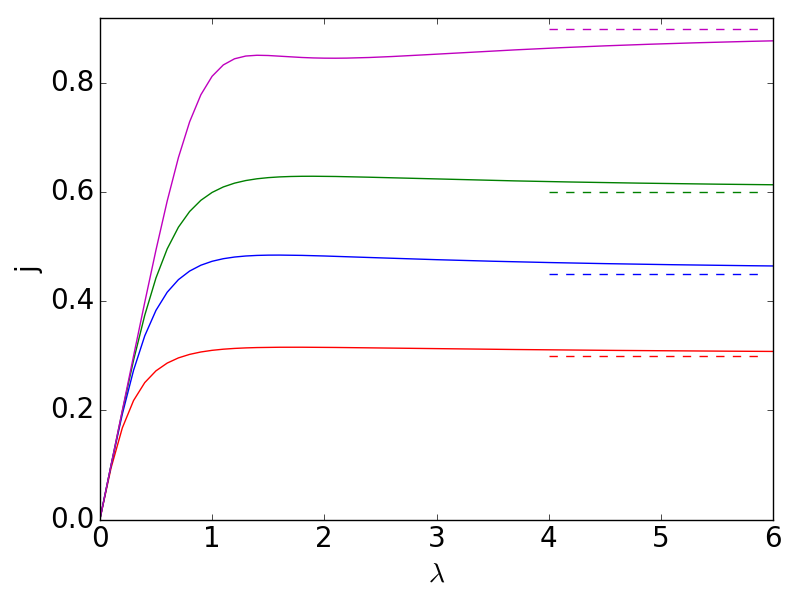}
\end{center}
\caption{Stationary flux $j$ as a function of $\lambda$ for $N=2,3,4,6$ (from bottom to top) with $\mu=1$ and for two values of $\mu^*$: (left) $\mu^*=1/2$ and (right)$\mu^*=0.1$. The dashed lines show the limiting value, $N\mu^*$.}\label{fig:flux}
\end{figure}
Figure \ref{fig:flux} shows the stationary flux versus $\lambda$ for the model with $N=2,3,4,6$ (from bottom to top). For $\mu^*=0.5$, at low intensity $\lambda$, the flux  increases linearly with $\lambda$ since blockage events are improbable, and 
finality saturates  towards  the limit, but the flux is a monotonically increasing function of $\lambda$. 
The dashed lines correspond to the limit $j=N\mu^*$.   For $\mu^*=0.1$, $j$ displays a maximum for a finite value of $\lambda$ for $N=2,3,4$, but not for $N=6$, where the limit is reached from below. One can show the critical value of $\mu^*$ for which the asymptotic behavior changes is  $\mu^*/\mu=0.25,0.22,0.187,0.139$ for $N=2,3,4,6$, respectively. 	 

To solve Eq.(\ref{eq:scm}), one notes that this process belongs to the class of circular Markov chains\cite{adan1997,Adan2002} for which a solution can be obtained for the stationary state.
After some calculations, the probability that the channel is blocked $\pi(1,0)$ is found as
\begin{equation}
 \pi(1,0)=\frac{1}{1+\frac{\mu^*}{\lambda}\sum_{i=1}^N (N+1-i)\left(\frac{\mu}{\lambda}\right)^{i-1}},
\end{equation}
and by using Eq.(\ref{eq:fluxs}), the stationary flux of exiting particles $j$ reads
\begin{equation}
 j=\frac{\mu^*\sum_{i=1}^N (N+1-i)\left(\frac{\mu}{\lambda}\right)^{i-1}}             	
 {1+\frac{\mu^*}{\lambda}\sum_{i=1}^N (N+1-i)\left(\frac{\mu}{\lambda}\right)^{i-1}}.
\end{equation}
The asymptotic regimes can be easily analyzed.
When $\lambda\rightarrow \infty$, the flux $j$ 
\begin{equation}\label{eq:fluxasumpt}
 j\simeq N\mu^*+\frac{\mu^*}{\lambda}((N-1)\mu-N^2\mu^*).
\end{equation}
The leading term corresponds to an alternation of  open (empty) and closed (blocked) states. The former is of (infinitesimally) short duration where no particle exit and the latter is followed by the release of $N$ blocked particles.  
This explains why the entrance flux $j$ is independent of $\lambda$ at large intensity. 
The second term of the asymptotic expansion, Eq.\ref{eq:fluxasumpt}, shows  that the limit is approached from below when $\mu^*<\frac{N-1}{N^2}\mu$ and from above when $\mu^*>\frac{N-1}{N^2}\mu$. This implies that, in the later case, the flux $j$ displays a maximum at a finite value of $\lambda$, whereas $j$ is a monotonically increasing function of $\lambda$ in the other case (A similar behavior is observed in a similar model see Ref.\cite{Page2018}).

At small  $\lambda$, $j$ is given by
\begin{equation}\label{eq:fluxsmall}
 j\simeq\lambda-\frac{\lambda^{N+1}}{\mu^{N-1}\mu^*}.
\end{equation}
The leading term of this  expansion expresses that all particles exit the channel without blockage and the sub-leading term corresponds to a decrement that becomes very small as the threshold $N$ increases.

\section{Multi-channel models}
\label{sec:multiple}
Because the number of probabilities necessary to define the model increases rapidly with the number of channels,
we   propose a graphical method which enables the enumeration of all possible events.

For $N=2$, the  state probabilities are given by  $\pi(i,j,t)$ where $i$ is an index counting 
the number of blocked channels and $j$ is an index for the number total of non-blocking particles. From these definitions, 
$i$ goes from $0$ to $N_c$ and $j$ from $0$ to $N_c-i$ and the total number of state probabilities, $N_p$ is
equal to $(N_c+1)(N_c+2)/2$. This gives $N_p=3,6,10,15$ for $N_c=1,2,3,4$, respectively. 

\begin{figure}[t!]
\begin{center}
\includegraphics[width=6cm]{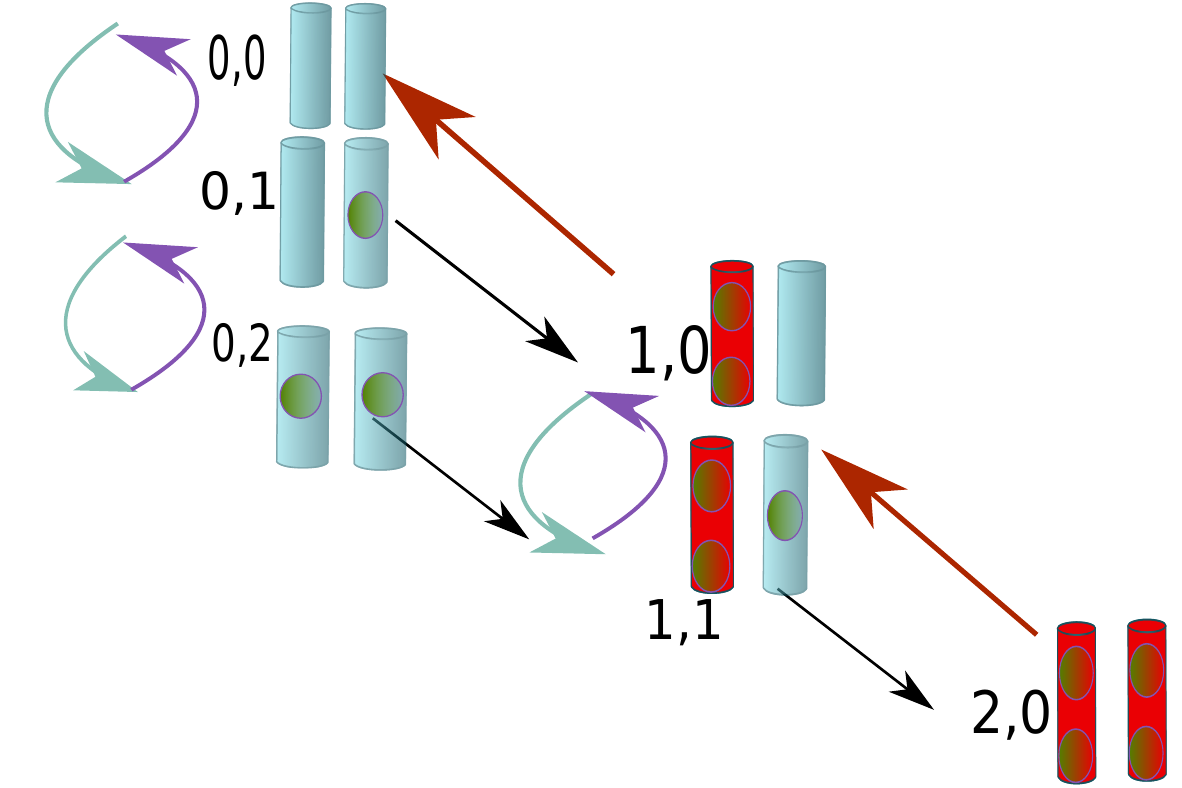}\includegraphics[width=6cm]{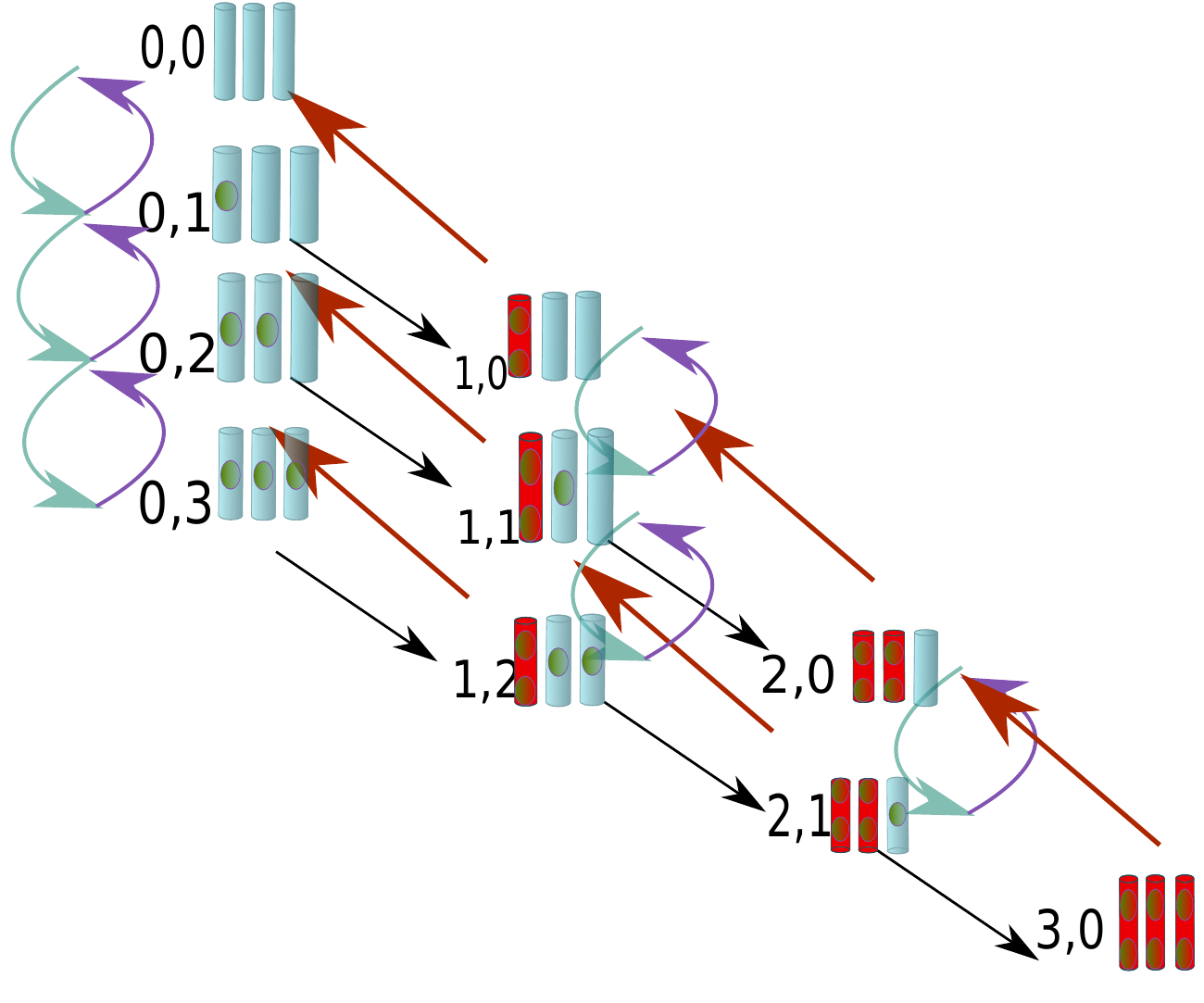}
\end{center}
\caption{State diagrams for the multi-channel models:  The left and right indices correspond to the number of blocked channels and the number
of channels with $1$ particle, respectively. Arrows represent all possible transitions between states: Red and blue curved arrows correspond to a transition between open states by the gain/loss of a particle, respectively. The black arrows correspond to a channel blockage due to a particle entrance and  the red arrows correspond to a channel release. (left) $N_c=2$ (right) $N_c=3$.
}\label{fig:diag}
\end{figure}

Figure \ref{fig:diag} displays state diagrams for  $N_c=2$ (left) and $N_c=3$ (right). Each state is labeled with the indices $(i,j)$ defined above. Transitions between states with open channels are shown by curved arrows and other arrows correspond to either a transition from an open to a blocked channel (black arrows) or transition between a blocked channel to an open channel (mauve arrows).

\subsection{$N_c=2$}
The kinetic equations of the model are given by the matrix differential equation
\begin{equation}\label{eq:eqdif2}
 \frac{dP(t)}{dt}=P(t).Q_{6},
\end{equation}
where $P(t)$ is the state vector with $6$ components,
$P(t)=(\pi(0,0,t),\pi(0,1,t)$, $\pi(0,2,t),\pi(1,0,t)$, $\pi(1,1,t), \pi(2,0,t))$
and the transition probability matrix $Q_{6}$
is given by

\begin{equation}
 Q_{6}=\left(
 \begin{array}{llllll}
 -\Lambda&\Lambda&0&0&0&0\\
 \mu&\scriptstyle-(\Lambda+\mu)&\frac{\Lambda}{2}&\frac{\Lambda}{2}&0&0\\
 0&2\mu&\scriptstyle-(\Lambda+2\mu)&0&\Lambda&0\\
 \mu^*&0&0&\scriptstyle-(\Lambda+\mu^*)&\Lambda&0\\
 0&\mu^*&0&\mu&\scriptstyle-(\Lambda+\mu^*+\mu)&\Lambda\\
 0&0&0&-2\mu^*&0&-2\mu^*
\end{array}
\right).
\end{equation}
The non-zero coefficients of the matrix correspond to the different arrows of Fig.\ref{fig:diag}. 
Let us consider some of the terms in detail.  The time evolution of $\pi(0,1,t)$ has three gain terms and two loss terms: 
the former correspond to the entry of one particle 
in an empty system, the exit of a particle from a system where each channel contains one  particle
(which explains the factor $2$), and the release of a blockage from a system with one blocked channel and a channel with one particle. 
The loss terms correspond to the entrance or the exit of a particle for a system in the state $(0,1)$. 
As a second example, the time evolution of $\pi(0,2,t)$ has one gain term associated with the entry of a particle in an empty channel
with the other channel containing one particle (which explains the $1/2$ factor), and two loss terms associated with the entrance of 
a new particle and the exit of one particle (the factor $2$ comes from the fact that a particle can exit from either channel).

The throughput of the two channel system is given by
\begin{equation}\label{eq:fluxnc2}
\fl J=\mu[\pi(0,1,t)+2\pi(0,2,t)]+2\mu^*\pi(1,0,t)+(\mu+2\mu^*)\pi(1,1,t)+4\mu^*p(2,0,t),
\end{equation}
where the first term corresponds to the exit of a particle from a system with no blocked channels.  The second term corresponds to a blockage release (the factor $2$ is the number of particles in the blocked channel). The third term contains two contributions: a particle exit from the open channel and a blockage release. The last term corresponds to the release of two particles from one of the two blocked channels.

In the stationary state, 
\begin{equation}\label{eq:fluxjnc2}
 J=\lambda(1-\pi(2,0)),
\end{equation}
which corresponds to the entrance flux times the probability that at least one of the channels is open.
One easily obtains the $\pi(2,0)$ 
\begin{equation}
 \pi(2,0)=\frac{2\Lambda^5+(2\mu+\mu^*)\Lambda^4}{\Delta},
\end{equation}
with
\begin{eqnarray}
 \Delta&=2\Lambda^5+(2\mu+9\mu^*)\Lambda^4+4\mu^*(3\mu+5\mu^*)\Lambda^3+ 2\mu^*(2\mu^2+21\mu\mu^*+4\mu^{*2})\Lambda^2
 \nonumber\\&+16\mu\mu^{*2}(2\mu+\mu^*)\Lambda+8(\mu\mu^*)^2(\mu+\mu^*).
\end{eqnarray}
By using Eq. (\ref{eq:fluxnc2}), one obtains the stationary throughput $J$. Some examples are shown in Fig. \ref{fig:fluxmc}.

At small $\lambda$, $J$ behaves as
\begin{equation}
 J\simeq \Lambda-\frac{2\mu+\mu^*}{8(\mu\mu^*)^2(\mu+\mu^*)}\Lambda^5.
\end{equation}
Comparing to the single channel model (with $N=2$) (Eq.\ref{eq:fluxsmall}), where the first term has a $\lambda^3$ dependence,  the first term has now
a $\lambda^5$ dependence, which corresponds smaller probability of a full blockage of the system.
At large $\Lambda$, one obtains
\begin{equation}
 J\simeq 4\mu^*-\frac{2\mu^*(\mu-4\mu^*)}{\Lambda}.
\end{equation}
As expected, one recovers that in this limit the throughput is the result of blockage release only, because incoming particles cannot cross the channel without causing a blockage. We note that the limit is approached from below for $\mu^*<\mu/4$ and from above for $\mu^*>\mu/4$. The latter case results in a maximum of the flux for a finite value of $\lambda$. See Fig. \ref{fig:fluxmc}. This change of behavior for the two-channel model at $\mu^*=\mu/4$ is the same as in the one channel model.

\subsection{$N_c\geq3$}
It is possible to derive the time evolution for a model with a number of channels larger than $2$, even if the calculation become rapidly cumbersome.
The time evolution of the process with $N_c=3$ is given by a system of differential equations,  Eq.(\ref{eq:eqdif2}) where the state vector $P(t)$ is given as a $10$-component vector, $P(t)=(\pi(0,0,t), \pi(0,1,t)$, $\pi(0,2,t), \pi(0,3,t), \pi(1,0,t), \pi(1,1,t), \pi(1,2,t), \pi(2,0,t), \pi(2,1,t), 
\pi(3,0,t))$  and the matrix $Q_{10}$ is given by

\begin{figure}[t!]
\begin{center}
\includegraphics[width=7.0cm]{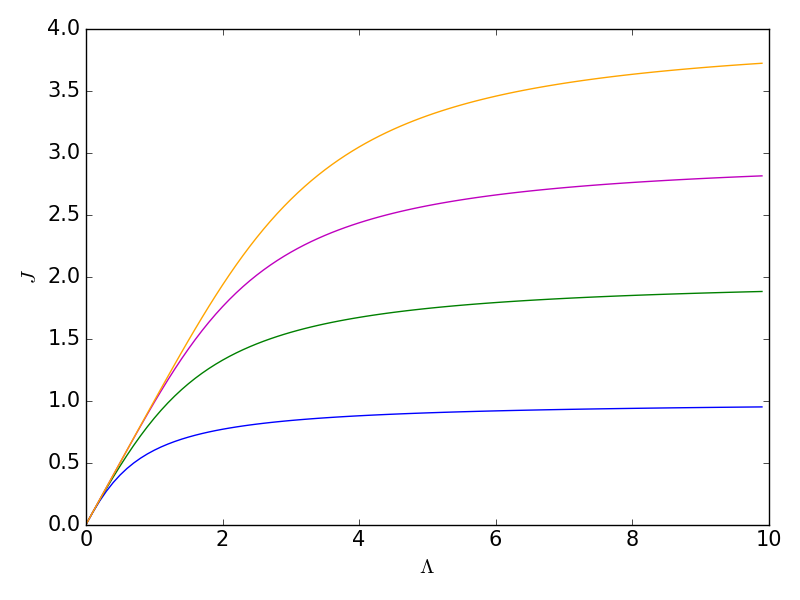}\includegraphics[width=7.0cm]{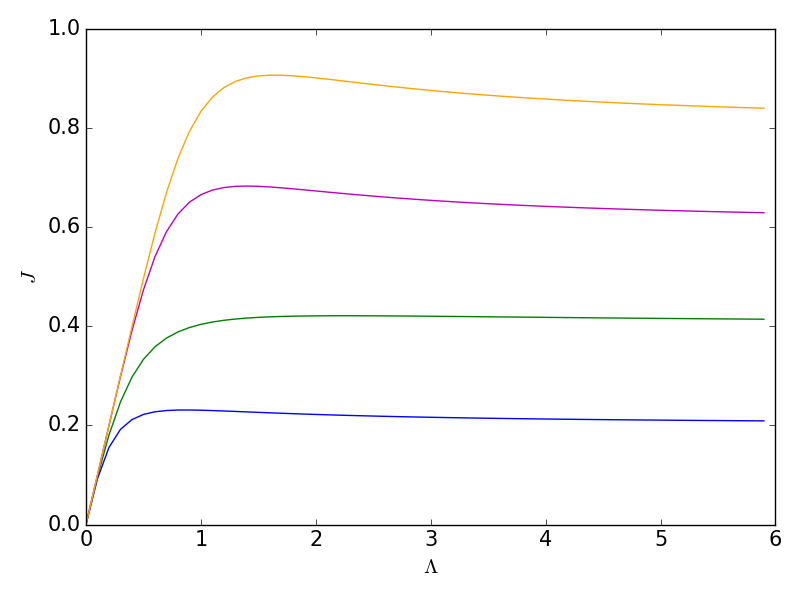}
\end{center}
\caption{Stationary flux $J$ as a function of $\Lambda$ for $N_c=1,2,3,4$ (from bottom to top) with $N=2$, $\mu=1$ and for two values of $\mu^*$: (left) $\mu^*=1/2$ and (right) $\mu^*=0.1$.
}\label{fig:fluxmc}
\end{figure}

\begin{equation}
 \fl\left(\scriptstyle
 \begin{array}{llllllllll}
\scriptstyle -\Lambda&\scriptstyle\Lambda&0&0&0&0&0&0&0&0\\ 
 \scriptstyle\mu&\scriptstyle-(\Lambda+\mu)&\frac{2\Lambda}{3}&0&\frac{\Lambda}{3}&0&0&0&0&0\\
 0&\scriptstyle 2\mu&\scriptstyle-(\Lambda+2\mu)&\frac{\Lambda}{3}&0&\frac{2\Lambda}{3}&0&0&0&0\\
 0&0&\scriptstyle 3\mu&\scriptstyle-(\Lambda+3\mu)&0&0&\scriptstyle\Lambda&0&0&0\\
 \scriptstyle\mu^*&0&0&0&\scriptstyle-(\Lambda+\mu^*)&\Lambda&0&0&0&0\\
 0&\scriptstyle\mu^*&0&0&\mu&\scriptstyle-(\Lambda+\mu^*+\mu)&\frac{\Lambda}{2}&\frac{\Lambda}{2}&0&0\\
 0&0&\scriptstyle\mu^*&0&0&\scriptstyle2\mu&\scriptstyle-(\Lambda+\mu^*+2\mu)&0&\Lambda&0\\
 0&0&0&0&\scriptstyle 2\mu^*&0&0&\scriptstyle-(\Lambda+2\mu^*)&\Lambda&0\\
 0&0&0&0 &0&\scriptstyle 2\mu^*&0&\scriptstyle \mu&\scriptstyle-(\Lambda+2\mu^*+\mu)&\scriptstyle \Lambda\\
 0&0&0&0&0&0&0&\scriptstyle 3\mu^*&0&-3\mu^*
\end{array}
\right)
\end{equation}

The exact solution is too lengthy to be displayed, but we focus on some partial results.
The stationary throughput is given by
\begin{equation}
 J=\lambda(1-\pi(3,0)).
\end{equation}
At low $\lambda$, one obtains
\begin{equation}
J\simeq \lambda -\frac{18\mu^3+39\mu^2\mu^µ+22\mu\mu^{*2}+4\mu^{*3}}{324(\mu\mu^*)^3(\mu+2\mu^*)(\mu+\mu^*)(2\mu+\mu^*)}\lambda^7.
\end{equation}
We have also solved the model for $N_c=4$ with $15$ probabilities, but the expressions are very lengthy. We present the results graphically in Figs.\ref{fig:fluxmc} and \ref{fig:diffluxmc}.

For general $N_c$ we conjecture that the small $\lambda$ expansion is
\begin{equation}
 J\simeq\lambda-\lambda^{2N_c+1}f(\mu,\mu^*),
\end{equation}
where $f(\mu, \mu^*)$ is a positive function of $\mu$ and $\mu^*$,
and the  $\lambda\rightarrow \infty$ limit is
\begin{equation}
 J=2N_c\mu^*+\frac{N_c\mu^*(\mu-4\mu^*)}{\lambda}.
\end{equation}

Figure \ref{fig:fluxmc} shows the stationary flux $J$ as a function of $\Lambda$ for $N_c=1,2,3,4$. As discussed above, $J$ is a monotonically increasing function when $\mu^*=0.5$, whereas $J$ displays  a maximum for $\mu^*=0.1$, a result that is independent of the number of channels $N_c$.

\section{Optimized transport}

\begin{figure}[t!]
\begin{center}
\includegraphics[width=7.0cm]{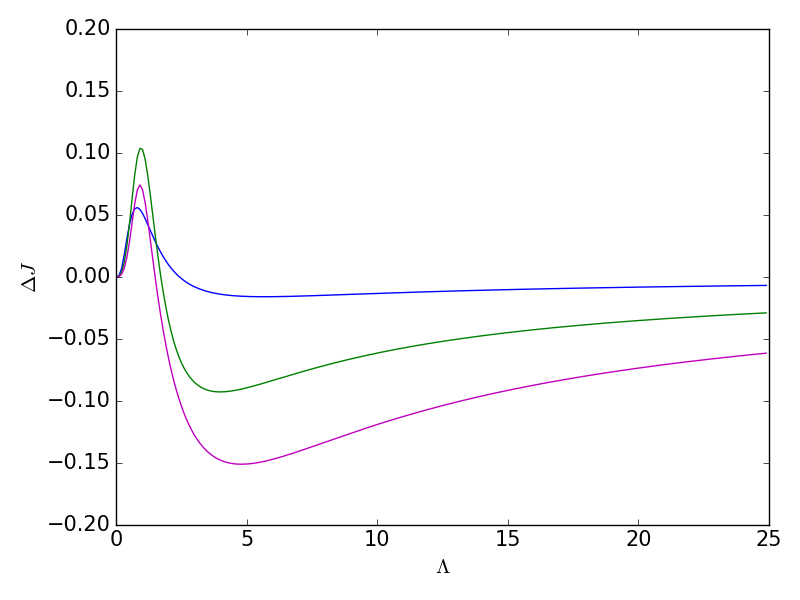}\includegraphics[width=7.0cm]{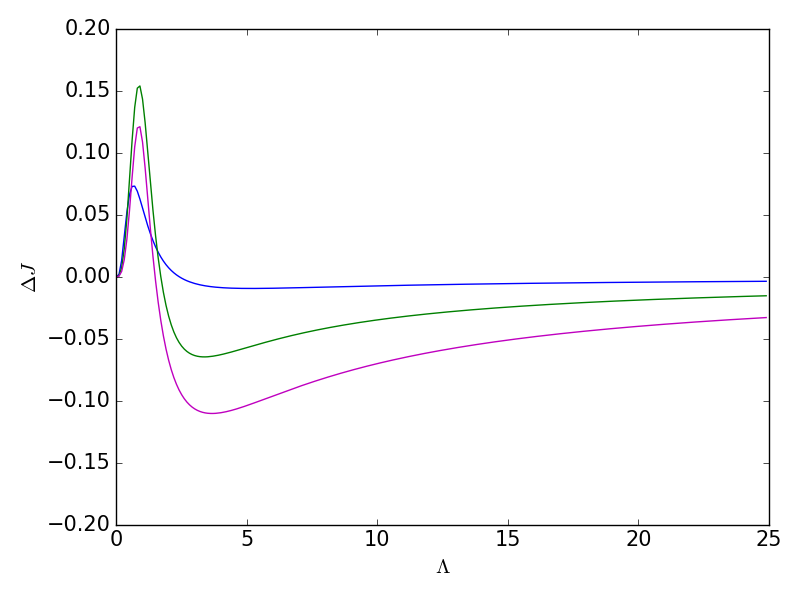}
\end{center}
\caption{ Flux difference $\Delta J$ as a function of $\Lambda$ for $N=8,6,4$ and $n=2,3,4$ with $\mu=1$ and $\mu^*=1/2$ (left) and $\mu^*=0.1$ (right)
}\label{fig:difflux}
\end{figure}

Here we examine different scenarios for conveying a particulate flux of given intensity $\Lambda$. The first scenario compares a single ``robust'' channel with a threshold equal to $N_1$ with a set of ``fragile'' $n$ identical channels each with a threshold of $N_1/n$ where the intensity is equally distributed over all channels. The second compares a system of $N_c$ coupled channels, i.e. the total intensity is equally distributed over all open channels, each with threshold $N=2$ with $N_c$ independent channels each with threshold $N=2$. In both cases we seek to determine which of the two configurations optimizes the steady state throughput.

{\bf One ``robust" channel versus several ``fragile" channels.} At low intensity, both configurations present few blockage events and the throughputs are the same and equal to $\Lambda$. At large intensity, the throughput of the ``robust'' channel is equal to $N_1\mu^*$ and since each ``fragile'' channel has a throughput equal to $(N_1/n)\mu^*$, the total outgoing flux is also equal to $N_1\mu^*$.

Fig.\ref{fig:difflux} displays the difference of throughput $\Delta J=J_{2n}-nJ_{2}$ between the stationary flux of the ``robust'' channel $J_{2n}$ and the sum of each individual flux of the ``fragile'' channels, $J_{2}$, as a function of $\Lambda$ for two values of $\mu^*$, ($\mu=1$) and for $n=2,3,4$. 
One  observes that the ``robust'' channel is more efficient at low intensity. The throughput difference reaches a maximum for a finite value of $\Lambda$, passes through zero  before attaining a negative minimum, corresponding the maximum of efficiency for the set of ``fragile'' channels. 
By using the first order expansion of the flux, one can explain easily that $\Delta J$ is not a flat function of $\Lambda$:
At low intensity $\Lambda$,
one has
\begin{equation}
 \Delta J\simeq\frac{\Lambda^3}{\mu^*\mu^2}\left[n-\left(\frac{\Lambda}{\mu}\right)^{2(n-1)}\right],
\end{equation}
which is positive for $n>1$.
Conversely at high intensity
\begin{equation}
 \Delta J=-\frac{(n-1)^2\mu^*\mu}{\Lambda},
\end{equation}
which is always negative.

\begin{figure}[t!]
\begin{center}
\includegraphics[width=7.0cm]{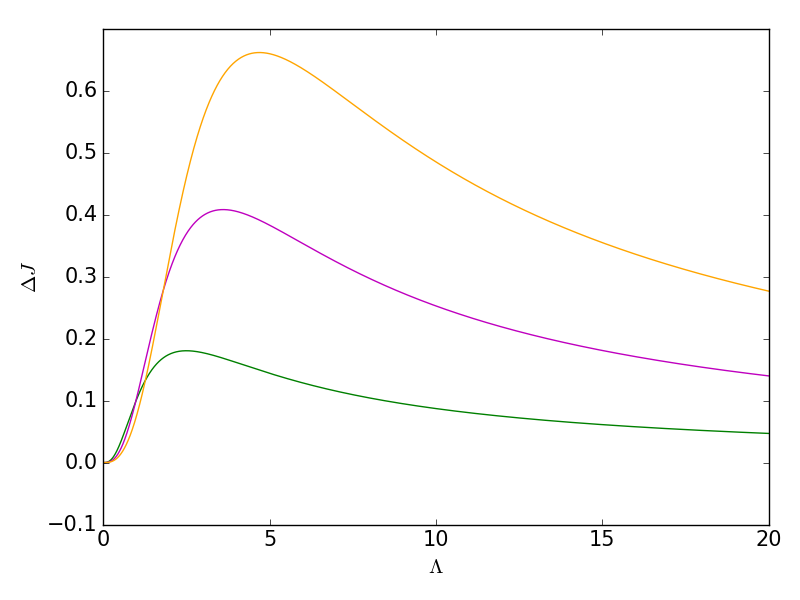}\includegraphics[width=7.0cm]{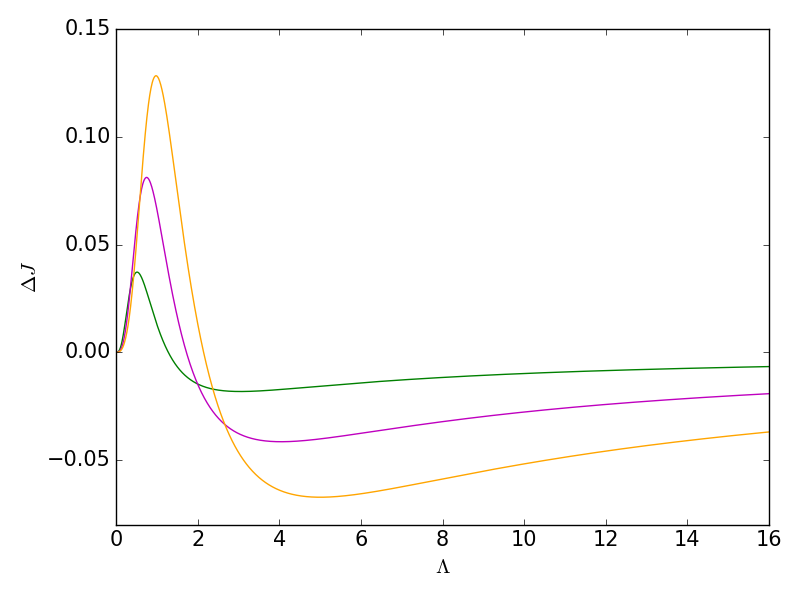}
\end{center}
\caption{Flux difference $\Delta J$ as a function of $\Lambda$ for $N=2$ and $N_c=2,3,4$ (lower and upper curves, respectively)  with $\mu=1$ and $\mu^*=1/2$ (left) and $\mu^*=0.1$ (right).
}\label{fig:diffluxmc}
\end{figure}

{\bf $N_c$ coupled channels with $N=2$ versus $N_c$ independent channels with $N=2$.} 
At low intensity, there are few blockages in either system and the throughput is $\Lambda$ in both. 
At large intensity, the throughput of the independent channels and the coupled correlated channels are both equal to $2N_c\mu^*$.

 Fig.\ref{fig:diffluxmc} shows the difference of throughput, $\Delta J(\Lambda)=J_{N_c}(\Lambda)-N_cJ_{2}(\Lambda/N_c)$,  between  the system of the $N_c$ coupled channels with $J_{N_c}$ and 
 the sum of the flux of each independent channels flux (with a threshold equal to $2$) $N_cJ_{2}$ as a function of $\Lambda$ for two values of $\mu^*$, ($\mu=1$) and for $N_c=2,3$. 

 
If the deblockage rate is sufficiently large, $\mu^*>0.25$, the independent channels  always convey the flux less efficiently than the coupled channels. 
If $\mu^*<0.25$ the behavior is similar to the first scenario: $\Delta J$ reaches a maximum for a finite value of $\Lambda$ and for higher intensity,  
the coupled channels are less efficient and $\Delta J$ reaches a minimum. At very large intensity $\Lambda$, both models converge to the same limit as expected.

This behavior can be understood by examining the limiting behavior of $\Delta J$. For small $\Lambda$ one has (for $N_c>1$)
\begin{equation}
 \Delta J\simeq\frac{\Lambda^3}{N_c^2\mu^*\mu^2},
\end{equation}
which is always positive, while at high intensity
\begin{equation}
 \Delta J\simeq-\frac{N_c(N_c-1)\mu^*}{\Lambda}(\mu -4\mu^*),
\end{equation}
which is negative if $\mu^*/\mu<0.25$ and positive otherwise. Coupled channels are always more efficient at low intensity and also at high intensity if the deblockage rate is sufficiently
high. If, however, $\mu^*<\mu/4$ the coupled channels convey the flux less efficiently due to an accelerating cascade of blockages that is 
reminiscent of the irreversible model \cite{Barre2015b}. 

\section{Summary}

We have presented a stochastic model of blockage in a channel bundle consisting of $N_c$ individual channels. A particulate flux enters the system according to a Poisson process of intensity $\Lambda$. 
Particles exit open channels at a rate $\mu$.  An individual channel is blocked if $N$ particles are simultaneously present in it. In this case, the flux that would have entered it is evenly distributed over the remaining open channels. A channel remains blocked for an exponentially distributed time with rate $\mu^*$. If all channels are blocked, the entering flux is rejected. We have provided a framework to obtain both the time-dependent and steady state properties and have presented explicit results for the steady state throughput for $N_c=2,3$. 

We used the theory to compare different methods for transporting a particulate flux of given intensity. A single robust channel of high capacity is more efficient than several fragile channels at low intensity, but the reverse is true at higher values of $\Lambda$. 
We also compared $N_c$ coupled channels with $N=2$ with the uncoupled version. The coupled channels always have a higher throughput than the independent channels if $\mu^*/\mu>0.25$.  For $\mu^*/\mu<0.25$ the coupled channels are more efficient at low intensity, but at higher intensities the order reverses. 
It will be interesting to see if this effect is still present when the dynamics is non-Markovian.

\section*{References}


\providecommand{\newblock}{}

\end{document}